\documentclass[12pt]{iopart}
\usepackage{epsf}
\begin{document}
\title{Secure quantum channels with correlated twin laser beams}
\author{Constantin V. Usenko\dag\ and Vladyslav C. Usenko\ddag
\footnote[3]{To whom correspondence should be addressed (usenko@univ.kiev.ua)} }

\address{\dag\ National Shevchenko University of Kyiv, Department of Theoretical Physics, Kyiv, Ukraine}

\address{\ddag\ Institute of Physics of National Academy of Science, Kyiv, Ukraine}

\begin{abstract}
This work is the development and analysis of the recently proposed quantum cryptographic protocol,
based on the use of the two-mode coherently correlated states. The protocol is supplied with the
cryptographic control procedures. The quantum noise influence on the channel error properties is examined.
State detection features are proposed.
\end{abstract}

\section{Introduction.}

The goals of quantum cryptography and secure quantum communications 
\cite{qc1, qc2, qc3} can be achieved using various protocols, which were 
developed and realized \cite{entprot, fourexp} in the past years on the basis 
of the quantum entanglement \cite{ent1, entprot} 
of weak beams and the single \cite{single1, single2} or few photon states \cite{four}, 
mostly by means of adjusting and detecting their polarization angles \cite{polar}. 

Another method, based on the usage of the two-mode coherently correlated (TMCC) beams
was proposed recently \cite{tmcc}. In this case the secure cryptographic key is generated
by the laser shot noise and duplicated through the quantum channel. Unlike the single or few photon schemes, 
which require large numbers of transmission reiterations to obtain the statistically significant 
results, the TMCC beam can be intensive enough to make each single measurement 
statistically significant and thus to use single impulse for each 
piece of information, and remain cryptographically steady.

In this work we analyse the error properties of the secure quantum channels, based on the
TMCC-beams and propose some additions to the TMCC-based cryptographic protocol.

The two-mode coherently correlated state is the way we refer to the 
generalized coherent state in the meaning by Perelomov \cite{per}. Such state can 
be described by its presentation through series by Fock states:

\begin{equation}
\label{eq:tmcc}
\left| \lambda \right\rangle =\frac{1}{\sqrt{I_0\left(2\left|\lambda\right|\right)}}\sum_{n=0}^\infty {\frac{\lambda ^n}{n!}\left| {nn} 
\right\rangle } 
\end{equation}

Here we use the designation $\left| {nn} \right\rangle = \left| n 
\right\rangle _1 \otimes \left| n \right\rangle _2 $, where $\left| n 
\right\rangle _1 $ and $\left| n \right\rangle _2 $stand for the states of 
the $1^{st}$ and $2^{nd}$ modes accordingly, represented by their photon 
numbers. The states (\ref{eq:tmcc}) are not the eigenstates for each of the operators 
separately, but are the eigenstates for the product of annihilation 
operators: 
\begin{equation}
	a_1 a_2 \left| \lambda \right\rangle = \lambda \left| \lambda 
	\right\rangle . 
\end{equation}
Such states can also be obtained from the zero state:

\begin{equation}
\label{eq:tmccground}
\left| \lambda \right\rangle = \frac{1}{\sqrt{I_0\left(2\left|\lambda\right|\right)}}I_0 (\lambda a_1^ + a_2^ + )\left| 0 
\right\rangle 
\end{equation}

In this work we assume that two laser beams, which are 
propagating independently from each other, correspond to the two modes of 
the TMCC state. States of beams are mutually correlated (surely, the TMCC 
state can also be represented in another way, for example, as a beam 
consisting of two correlated polarizations).

\section{Beam measurement}

Let's examine any of the two TMCC beams separately. 
The intensity of the beam's radiation, registered by an observer is proportional to the 
mean of the $N = a^ + a$ operator, which is the number of the photons 
in the corresponding mode. 
The mean observable values, which characterize the results of the 
measurements of the beam are:

\begin{equation}
\label{eq:meann}
\left\langle {N } \right\rangle = \left\langle \lambda \right|a^ 
+ a \left| \lambda \right\rangle , \left\langle {N^2 } \right\rangle = 
\left\langle \lambda \right|a^ + a a^ + a \left| 
\lambda \right\rangle 
\end{equation}

These characteristics are squared in field, and thus their mean values don't turn to 
zero (this fact is not specific for the TMCC-states, because the usual non-correlated 
states and processes, like the heat propagation, show the same properties).

Assuming the state expression (\ref{eq:tmcc}) we obtain

\begin{equation}
\label{eq:meann2}
\langle {N } \rangle = \frac{1}
{I_0 ( 2| \lambda | )}\sum_{n=0}^\infty n \frac{| \lambda |^{2n}}{n!^2} , \langle {N^2 } \rangle = \frac{1}
{I_0 ( 2| \lambda | )}\sum_{n=0}^\infty n^2 \frac{| \lambda |^{2n}}{n!^2}
\end{equation}

The mean number of registered photons is

\begin{equation}
\langle N \rangle = \sum_{n=0}^\infty nP_n (\lambda)
\end{equation}

The probability of registering n photons depends on the intensity of a beam:

\begin{equation}
\label{eq:nphotprob}
P_n(\lambda) = \frac{1}{I_0 (2|\lambda|)}\frac{|\lambda|^{2n}}{n!^2} 
\end{equation}

An important feature of this distribution is the quick (proportional to $n!^2$) decreasing dependence of the registration probability
on the photon number. 

\begin{figure}[htbp]
\begin{center}
	\epsfbox{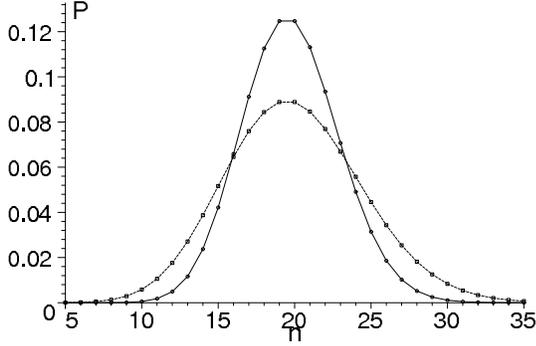}
	\caption{ Probability of different photon numbers registration distribution for the TMCC-beam (circles, solid line) and the analogous distribution for a usual coherent beam (squares, dotted line)}
	\label{plot_pn}
\end{center}
\end{figure}

This circumstance makes the experimental identification of the
TMCC-states quite convenient. The distribution of the probability of different photon 
numbers registration along with the analogous distribution for a usual coherent beam 
are given at the \fref{plot_pn}. One can see that there are significant differences 
for the TMCC and the Poisson beam distributions - the TMCC-beam distribution is relatively 
sharp and narrow.

Taking into account (\ref{eq:nphotprob}) the expressions for the mean and mean 
square values of the registered photon numbers (\ref{eq:meann2}) turn to:

\begin{equation}
\label{eq:meann3}
\langle N \rangle = \frac{|\lambda|^2 I_1(2|\lambda|)}{I_0 (2|\lambda|)} , 
\left\langle {N^2 } \right\rangle = \left| \lambda \right|^2
\end{equation}

The measurements have the statistical uncertainty, caused by quantum 
fluctuations. This uncertainty can be characterized 
by the corresponding dispersion:

\begin{equation}
\sigma ^2 = \langle {N^2} \rangle - \langle 
{N} \rangle ^2
\end{equation}

Taking into account (\ref{eq:meann3}) we get the following 
expression:

\begin{equation}
\label{eq:sigma}
\sigma ^2 = \left| \lambda \right|^2\left( {1 - \left( {\frac{I_1 
\left( {2\left| \lambda \right|} \right)}{I_0 \left( {2\left| \lambda 
\right|} \right)}} \right)^2} \right)
\end{equation}

\begin{figure}[htbp]
\begin{center}
	\epsfbox{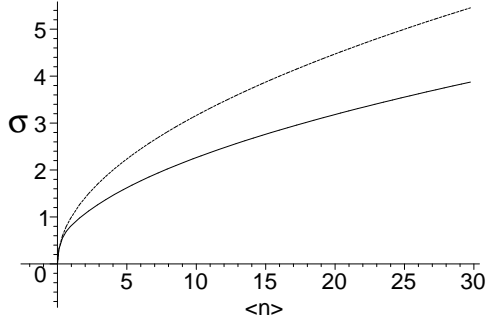}
	\caption{The dependencies of the measurement results uncertainty on the mean photon number for the TMCC-beam (solid line) and a usual coherent beam (dotted line)}
	\label{plot_dispers}
\end{center}
\end{figure}

The dependencies of the measurement results uncertainty on the mean photon number for
the TMCC-beam and a usual correlated beam are given at \fref{plot_dispers}. 
The difference between these dependencies can also be used for the TMCC-states identification.

\section{Communication via quantum channel}

\begin{figure}[htbp]
	\epsfbox{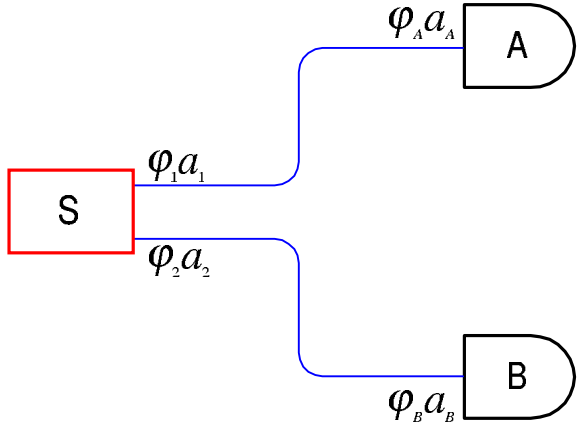}
	\caption{Quantum channel between two parties with a TMCC source}
	\label{scheme1}
\end{figure}

Let we have to establish a secure quantum channel between two parties (\fref{scheme1}). Alice has the laser on her side, which produces two beams in the TMCC 
state. The optical channel is organized in such a way, that Alice receives one 
of the modes, the first, for example, i.e. $\varphi _A \equiv \varphi _1 
$,$\varphi _A (x_A ,t_0 ) = 1$ , and Bob receives another one, i.e. $\varphi _B 
\equiv \varphi _2 $ ,$\varphi _B (x_B ,t_0 ) = 1$ at any moment of 
measurement $t_0 $, where $x_A $and $x_B $are Alice's and Bob's locations 
respectively. Accordingly, Alice cannot measure the Bob's beam and vice 
versa:$\varphi _B (x_A ,t_0 ) = 0$, $\varphi _A (x_B ,t_0 ) = 0$. At that 
the vector-potential of the TMCC-beam is:

\begin{equation}
A = \varphi _A^\ast (x,t)a_A^ + + \varphi _A (x,t)a_A + \varphi _B^\ast 
(x,t)a_B^ + + \varphi _B (x,t)a_B 
\end{equation}

Unlike the usual non-correlated coherent states, which
show their quasiclassical properties in the fact, that the mean value of a vector-potential of
a corresponding beam is not equal to 0, the mean value of a vector-potential of a TMCC-beam 
and any other characteristic, which is linear in field, turns to be equal to 0, because during 
the averaging by the 
1$^{st}$ mode, for example, the $a_1$ converts $\left| {n,n} \right\rangle $ to $\left| 
{n - 1,n} \right\rangle $, which is orthogonal to all the present state 
terms, so $\left\langle {\lambda _i } \right|a_i \left| {\lambda _i } 
\right\rangle = 0$. So the quasiclassical properties in their usual meaning are absent in 
the case of a TMCC-beam. But they become apparent in the non-zero value of the spatial correlation function,
which characterizes the interdependence of the results of measurements 
taken by Alice and Bob:

\begin{equation}
g_{AB} = < N_A N_B > - < N_A > < N_B > 
\end{equation}

It's useful to describe the channel quality by the relative correlation, 
which is

\begin{equation}
\rho _{AB} = \frac{ < N_A N_B > - < N_A > < N_B > }{\sigma _A \sigma _B }
\end{equation}

The main feature of the TMCC state is that the value $\rho _{AB} $ is 
exactly equal to 1, while in the case of non-correlated beams we would get 
$\rho _{AB} = 0$. This means that the measurements of the photon numbers, 
obtained by Alice and Bob, each with her/his own detector, not only show the same 
mean values, but
 even have the same deflection from the mean values.

The laser beam is the semi-classical radiation with well defined phase, but 
due to the uncertainty principle for the number of photons and the 
phase of the radiation, there is a large enough uncertainty in the photon 
numbers, this can be seen from the dispersion expression (\ref{eq:sigma}). Thus one can 
observe the noise, which is similar to the shot noise in an electron tube. 
In the TMCC radiation the characteristics of such noise for each of the 
modes are amazingly well correlated to each other. This fact 
enables the use of such radiation for generation of a random code, which 
will be equally good received by two mutually remote detectors.

\subsection{The protocol}

The following scheme can be used for the TMCC-based protocol. The laser is 
set up to produce the constant mean number of photons during the session and 
both parties know this number. At some moment Alice and Bob start the 
measurements. They detect the number of photons at unit time  by measuring 
the integrated intensity of the corresponding incoming beam. If the number 
of photons for the specific unit time is larger than the known expected mean 
(which is due to the shot noise), the next bit of the generated code is 
considered to have the value ``1''. If the measured number is less than 
the expected mean, the next bit is considered to be equal to ``0'':
 
\begin{equation}
\label{protocol}
B ={\left\{
\begin{array}{l}
\{n \leq [<N>]\} \rightarrow 0 \\ \{n > [<N>]\} \rightarrow 1
\end{array}\right.}
\end{equation}

Upon the receipt of a sufficient number of bits (the code), both Bob and Alice 
divide them in half, each obtaining two bit sequences (half-codes). Bob encodes one 
sequence with another, using the "eXclusive OR" logical operation (XOR, $B_i\bigoplus B_j$), 
and sends this encoded half-code to Alice using any public channel. 
Alice uses any of her half-codes to
decode the code she has got from Bob using the same XOR operation. She compares
the result of this operation with another of her half-codes. If
all the bits coincide, this means that Alice and Bob both have the same
code, which can be used as a cryptographic key for encoding their communication.
Otherwise they have to repeat the key generation and transfer procedure and check 
the channel for the possible eavesdropping if the procedure fails again.

The stability of the protocol against the basic beam splitting attacks was
examined in \cite{tmcc} and it was shown that any successful
attempt destroys the channel and cancels the key distribution session.
Besides the basic listening-in, Eve may carry out a more advanced
state cloning eavesdropping attack by detecting the overall photon
number in the Bob's mode for each upcoming bit and then re-emiting the
same number again, which requires Eve to have the same laser source on her site.
Though, due to the laser shot noise Eve can't be sure if she is
producing exactly the same number of photons, she can set up her
laser just to produce some mean photon number. In case she will
adjust her laser to produce the mean photon number, which is equal to
the current photon number measured in the Bob's mode for a next
bit, she may probably be successfull by repeating some bits. But in
this case she will change the Bob's measurement results ditribution.
This can be checked by expanding a protocol with a post-measurement
analysis of the measurement results, which can be done by both of
the trusted parties. It should consist of the comparison of an
obtained photon numbers distribution by the frequencies of their
detection with the expected one for the known constant mean photon number
value which is actually a task of comparing two numerical arrays
The difference between expected and obtained distributions will
reveal a state cloning attempt.

\subsection{Quantum channel error analysis}

Let the parties of the secret key transmission procedure are using the protocol 
described above, thus they estimate the value of the next bit by comparing the 
actual registered photon number to the average. The probability of detecting "0" 
bit value then is

\begin{equation}
P_{(0)} (\lambda) = \sum_{n=0}^{[\langle N \rangle]}P_n(\lambda)
\end{equation}

The noise is present in the channel and it may increase the number of the registered 
photons. We suppose that the noise is thermal and assume that 
it may, with some probability, cause an appearance of one and no more than one 
additional photon in any of the modes during the time of a bit detection. 
We will denote the probability of a noise photon detection as $\epsilon$ 
and refer to it as the noise factor.

We suppose that the channel is qualitative enough to transfer the impulse at the 
required distance without losing any single photon, thus errors are possible only 
due to the appearance of the noise photons.

An error, when Alice registers "0" bit value and Bob registers "1" may occur upon 
the joint realization of two events. The first is that Alice detects the maximum possible
number of these, corresponding to the "0" bit value, which is, according to the proposed protocol, 
equal to the integer part of $\langle n \rangle$. The second is that in addition to this number Bob 
detects the appearance of a noise photon. The opposite situation, when Alice gets the 
"1" bit value and Bob registers "0" is possible when the noise photon was detected 
by Alice, the probability of such error is the same. 

The probability of the realization of a state, which consists of the maximum possible 
for the "0" bit value photons and, at the same time, is detected as "0", is the 
relation between the corresponding probabilities:

\begin{equation}
P_{max(0)} = \frac{P_{[\langle N \rangle]}}{P_{(0)}} = \frac{P_{[\langle N \rangle]}}{\sum_{n=0}^{[\langle N \rangle]}P_n(\lambda)}
\end{equation}

We will refer to this probability as to the error factor.
So the probability of an error during the bit registration is equal to the product 
of the noise and error factors:

\begin{equation}
P_{err(0)}(\lambda)=\epsilon P_{max(0)} 
\end{equation}

One can easily see that upon the intensity increase the error factor becomes less and 
so the channel tends to a self-correction if the beam gets more intensive.

\section{Conclusions}

Correlated coherent states of the two-mode laser beam (TMCC states) show 
interesting properties, which can be used, in particular, for the tasks of 
the quantum communication and cryptography. 

The TMCC-beams can be identified due to the special form of the registration 
probabilities distribution for different photon numbers in the corresponding beam 
and the dependence of the dispersion on the mean photon numbers value.

On the one hand, each of the 
modes looks like a flow of the independent photons rather then a coherent 
beam, since mean values of the operators, which are linear in field, are 
equal to 0 for each mode separately. 

On the other hand, the strong 
correlation between the results of measurements for each of the modes takes 
place. This correlation shows itself in the fact that in each of the modes numbers 
of photons are the same and even the shot noise shows itself equally in the both 
modes. This enables the use of the TMCC state as the generator and carrier 
of random keys in a quantum channel which is stable against the eavesdropping \cite{tmcc}.

Thus, the TMCC-laser generates and transmits exactly the 2 copies of a 
random key. Unlike the single or two-photon schemes, which require large numbers of 
transmission reiterations to obtain the statistically significant 
results, the TMCC beam can be intensive enough to make each single measurement statistically 
significant and thus to use single impulse for each 
piece of information, and remain cryptographically steady. This allows to essentially increase 
the effective data transfer rate and distance. Analysis of the noise influence on
the channel properties shows that the channel tends to a self-correction upon the beam intensity increase.

\end{document}